\documentclass{PoS}

\title{Galaxy evolution begins at home: GALFA, EVLA, and GASKAP}

\ShortTitle{Galaxy evolution begins at home: GALFA, EVLA, and GASKAP}

\author{\speaker{Snezana Stanimirovic}%
         \thanks{A footnote may follow.}\\
        University of Wisconsin, Madison Department of Astronomy\\
        E-mail: \email{sstanimi@astro.wisc.edu}}

\abstract{ 

While studies of galaxy evolution generally focus on extensive HI surveys at large redshifts,
we argue in this paper that the understanding of {\it detailed} physical processes that drive HI evolution 
in galaxies is equally important. Specifically, we focus on three open questions regarding the 
very first step in the star-formation cycle in galaxies:
How much do galaxy halos flavor and tax the accretion flows that are postulated to bring
fresh star-formation fuel to galaxy disks?
What are the basic properties of the warm neutral gas, the 
progenitor of cold star-forming clouds?
And, what are the origin and level of interstellar inhomogeneities as seeding agents for 
molecule and star formation?
The very local Universe (The Milky Way and nearby galaxies) offers
an unparalleled high-resolution view for answering these questions
and the upcoming radio telescopes 
(e.g. EVLA, ASKAP, MeerKAT, ATA-256) promise great advances.

}

\FullConference{ISKAF2010 Science Meeting \\
                 June 10 -14 2010\\
                 Assen, the Netherlands}

\begin{document}

\def\degree{$^{\circ}$}
\def\kms{km~s$^{-1}$}
\def\ga{\mathrel{\hbox{\rlap{\hbox{\lower4pt\hbox{$\sim$}}}\hbox{$>$}}}}
\def\la{\mathrel{\hbox{\rlap{\hbox{\lower4pt\hbox{$\sim$}}}\hbox{$<$}}}}

\section{Introduction }

\begin{figure}
\begin{center}
\includegraphics[width=.8\textwidth]{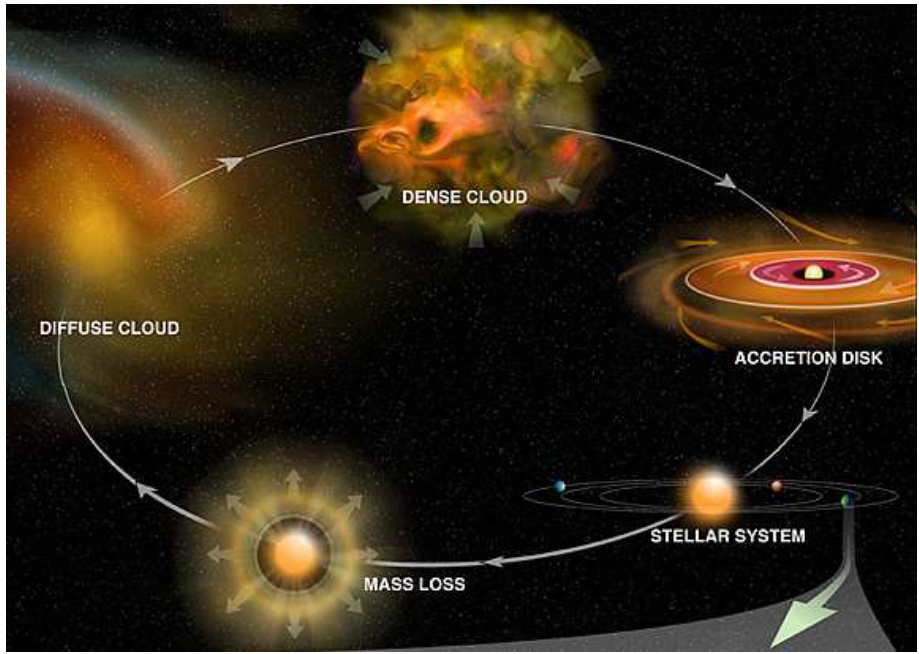}
\caption{Schematic summary of our current understanding of 
the star-formation cycle in galaxies. Credit: Bill Saxton (NRAO)}
\label{f:ism-cycle}
\end{center}
\end{figure}

In our current understanding of the star formation cycle in galaxies (illustrated in Figure~\ref{f:ism-cycle}),
diffuse  interstellar gas transforms first into dense cold clouds, 
which further fragment and produce stellar and planetary systems. 
Throughout their lifetime and particularly at the end, stars greatly 
affect the surrounding medium through stellar winds and supernovae  
stirring and structuring the diffuse gas and affecting the next generation of star formation.
While other components of this cycle have received significant attention,
the first step, or the conversion of diffuse interstellar gas into dense cold (molecular) clouds,
is largely unexplored despite the fact that it has long-reaching manifestations. 
For example, the outstanding ``missing satellite problem"  whereby
galaxy formation models over-predict the number of low-mass dark matter halos,
stems from our limited understanding of the physical processes 
(and their efficiencies) involved
in the molecular cloud and star formation (Putman et al. 2009).

Furthermore, recent cosmological simulations suggest an even higher complexity of the 
star-formation cycle by introducing one additional step: 
the accretion of the initial star formation  fuel (Kere\v{s} et al. 2005).
Even at the present time, a large fraction of the multi-phase diffuse gas in galaxies
is expected to be accreted from cosmic filaments and satellite galaxies 
enabling a healthy star formation rate. 
However, this process is not passive and the interplay between the inflowing material and the host galaxy
leaves a strong, multi-phase mark on the accreted gas.
In the case of a Milky Way (MW) type galaxy,
Brooks et al. (2009) show that about 60-70\% of the 
inflowing gas is shock-heated to near virial temperatures of $T\sim10^6$ K , 
while about 30-40\% is
accreted at lower temperatures of T$<$ few $\times$10$^5$~K 
through both cold accretion (from cosmic filaments) and accretion from previous
mergers and satellites (`clumpy' component).
The unshocked and `clumpy' components in particular play
an important role for building up the disk as the cold gas is delivered close to
the disk and goes on to form stars faster than the shocked gas, 
which must cool before supporting star formation.


The star formation cycle provides chemical and energy enrichment 
and directly affects how galaxies age and evolve with time.
Therefore, to make advances in our understanding of galaxy evolution,
we need to start with the necessary first step: the diffuse interstellar gas. 
We focus here on three scientific questions concerning the diffuse interstellar gas
where the upcoming radio telescopes promise great advances:
(i) what is the nature and fate of  the accreted star formation fuel;
(ii) what are the physical conditions required for cycling of interstellar phases; and
(ii) what is the origin and nature of interstellar inhomogeneities as seeds for molecule 
(and later star) formation?

\section{The Magellanic Stream as a template for {\it detailed} physics of accretion flows}

We are fortunate that the Magellanic Stream (MS) offers a nearby example of a gaseous
remnant from interactions between the Magellanic Clouds (MCs; 
the Large Magellanic Cloud, LMC, and the Small Magellanic Cloud, SMC) and the Milky Way (MW). This
feature, which extends in an arc nearly half way across the sky, offers a unique,
close-by laboratory to study physical processes of large-scale accretion flows in the MW halo.

The MS is a  huge ($>100$ degree long) starless neutral hydrogen (HI) structure
trailing behind the MCs. After decades of studies, numerous puzzles remain
regarding the formation and evolution of the MS gas (a recent summary is
provided in Stanimirovic et al. 2010).
Two most recent observational surprises are:
(i) the present-day MS is at least 40\% longer, and $\sim10$\% more massive, 
relative to the MS we knew about a few years ago (Braun \& Thilker 2004,
Stanimirovi{\'c} et al. 2008, Nidever et al. 2010);
and, (ii) the MS has a significant abundance of small-scale structure.

\begin{figure}
\begin{center}
\includegraphics[width=1.\textwidth]{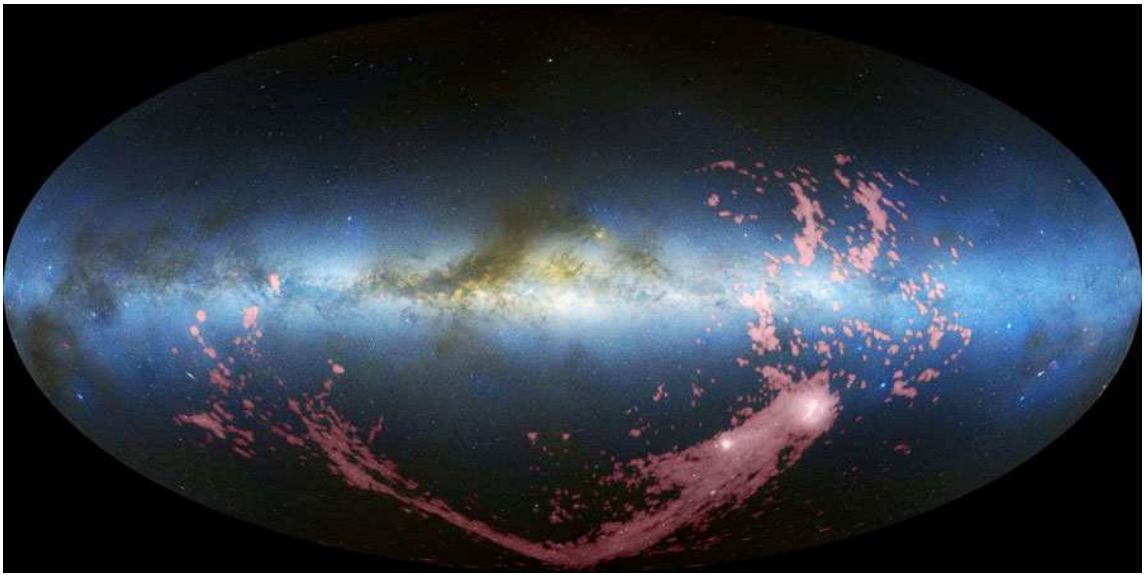}
\caption{The compilation of HI observations of the Magellanic Stream by David Nidever superimposed
on an all-sky image of the MW in visible light by Axel Mellinger. Credit: Astronomy picture of the day.
}
\label{f:stream}
\end{center}
\end{figure}

A compilation of various HI observations of the MS is shown in Figure~\ref{f:stream}
(Nidever et al. 2010, in preparation) and includes 
recent Arecibo HI observations from the GALFA-HI survey (Stanimirovic et al. 2006,
Peek \& Heiles 2008) focusing on the MS tip.
The GALFA-HI survey has been mapping the entire Arecibo sky at a velocity resolution
of 0.18 \kms and an angular resolution of 3.5 arcmin, operating mainly commensally 
with other surveys undertaken with the Arecibo L-Band Feed Array.
The complex small-scale morphology of the HI gas,
revealed for the first time down to an angular size of $\sim3.5$ arcmin,
indicates that processes are clearly at work in the MW halo 
on scales of tens of parsecs  and at a distance of $>60$ kpc.
These processes affect the MS's potential for star formation, the transfer of gas from the MS 
to the halo,  and also may provide additional drag affecting the global MS dynamics.
Although these processes play a crucial role for gas evolution of the MS
(Murray et al. 1993; Bland-Hawthorn et al. 2007; 
Heitsch \& Putman 2009), it is still not clear exactly how they operate, and on what timescales.
As global numerical simulations rarely have the resolution necessary to 
resolve such small scales, observational constraints of the 
effectiveness of various hydrodynamical instabilities are needed.

Analytical considerations of timescales (Mori \& Burkert 2001,
Quilis \& Moore 2001, Stanimirovic et al. 2008) as well as recent numerical advances
(Bland-Hawthorn et al. 2007, Heitsch \& Putman 2009), suggest
that thermal and Kelvin-Helmholz instabilities operate on timescales much shorter than the
MS formation time and hence must be important.  This, together with large-scale shearing
due to tidal interactions in the MW-MCs system, results in an expectation of a highly
turbulent environment. Yet, the physical properties of the MS gas revealed by 
the high-resolution observations indicate stability and longevity  of HI clouds. 
For example, cool HI cores subsonically moving within warmer HI envelopes 
have been found along the MS (Karberla \& Haud 2006, Stanimirovic et al. 2008). The 
coldest ($T\sim70$ K) pockets of HI have been recently revealed through
absorption observations by Matthews et al. (2009).
Diffuse ionized gas with $T\sim10^{4-4.5}$ K, further enveloping the HI component, 
has been studied with SiIII observations by Shull et al. (2009). An even hotter component with
$T\sim10^{5}$ K, observed through OVI absorption lines, has been
interpreted as tracing the interface between the cool MS gas and 
the hot MW halo ($T\sim10^{6}$ K).   From this work, a 
picture emerges of cold MS cores shielded from the $T\sim10^{6}$ K MW halo 
by many layers of a multi-phase warm gas.

Further, the direct comparison between the observed HI column density probability
density function (PDF) and the latest simulated distributions reveals significant differences.
As a demonstration, we use here data from the 
Bland-Hawthorn et al. (2007) shock-cascade model of the MS, which starts with
an initially clumpy HI distribution of the MS gas and
allows for strong interactions between the MS clouds and the MW halo.
As the MS clouds upstream experience gas ablation by the oncoming hot MW halo, 
the ablated gas is slowed down and  further
collides with the clouds downstream, resulting in shock ionization
of HI clouds. This shock-cascade model can explain measured H$\alpha$ intensities
along the MS (Bland-Hawthorn et al. 2007).
It also predicts large changes  in the HI distribution on timescales of 100-200 Myrs caused
by the ablation process. 

\begin{figure}
\begin{center}
\includegraphics[width=.8\textwidth]{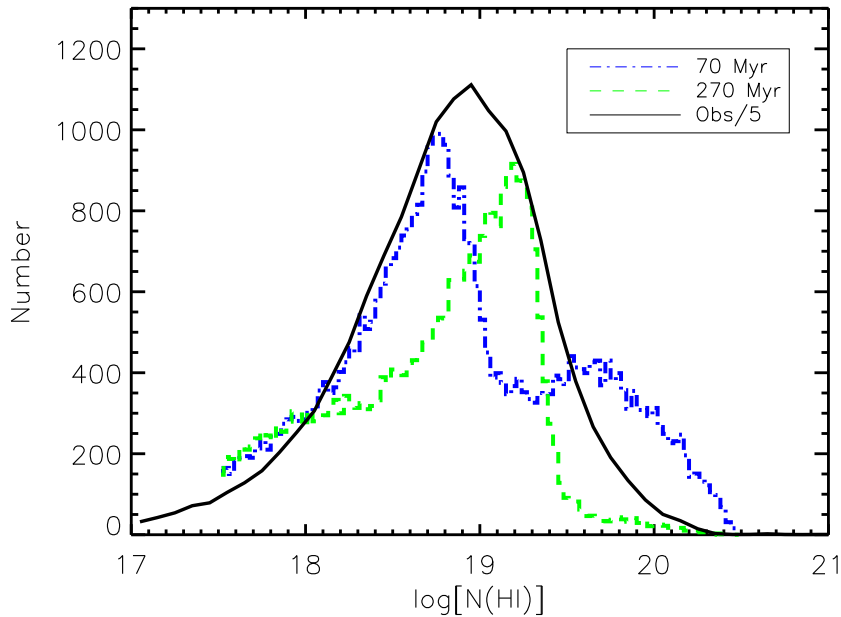}
\caption{Evolution of the HI column density probability density
function in the shock-cascade simulation by Bland-Hawthorn et al. (2007).
The blue dot-dashed and green dashed lines shows times stamps in the simulation
at 70 and 270 Myrs, respectively. The solid line shows the HI column density PDF
derived by using observations from Stanimirovic et al. (2008). 
The observed PDF was divided by 5
to account because observations sample five times larger projected area than the simulation.}
\label{f:joss}
\end{center}
\end{figure}

In Figure~\ref{f:joss} we compare the observed HI column density PDF for the MS tip
(from Stanimirovic et al. 2008)
with the  same quantity at two snap-shots in the Bland-Hawthorn et al.'s simulation:
70  and 270 Myrs after the initial exposure of the MS to the halo wind (shown
as dashed and dot-dashed lines in Figure~\ref{f:joss}). 
The large difference in the simulated
data after 200 Myrs  is clearly visible, and the later distribution is missing both
low- and high-density gas. 
However, the observed PDF is not similar to any of the
simulated PDFs. Contrary to a highly asymmetric
simulated N(HI) PDF, the observed PDF is highly symmetric and almost
Gaussian. It clearly contains more low- and high-density gas than the
end point of the simulation. As shown in Burkhart et al. (2010), subsonic turbulence
produces Gaussian column density distributions, while supersonic turbulence produces
highly skewed PDFs. This highlights the difference between observations and the simulation:
simulated distributions appear highly turbulent due to fast ablation processes.
As a result, the neutral gas is relatively quickly shredded and turned into an ionized warm drizzle, 
which eventually infalls onto the MW disk.

The structure of the boundary between clouds and the hot atmosphere of 
the MW is one factor that could slow down the rate of mass ablation in the MS through
heat conduction (e.g. Vieser \& Hensler 2007).  
To study these boundary regions along the MS, and also other tidal tails, 
we need large-area observations with both high angular resolution and excellent sensitivity. 
Several upcoming radio telescopes will provide exactly that (e.g. ASKAP, MeerKAT, ATA-256).
In particular, the Galactic spectral line survey with the Australian Square Kilometre Array Pathfinder
(GASKAP, Dickey et al. 2010), one of several survey science projects accepted for ASKAP, will
image the MS at an angular resolution of $\sim1$ arcmin.
We will be able to study the ``aging'' processes of the HI gas 
injected into the vicinities of galaxies by interactions or other cosmologically
related processes.


\section{What are the physical conditions required for cycling of interstellar phases?}

\begin{figure}
\begin{center}
\includegraphics[width=.8\textwidth]{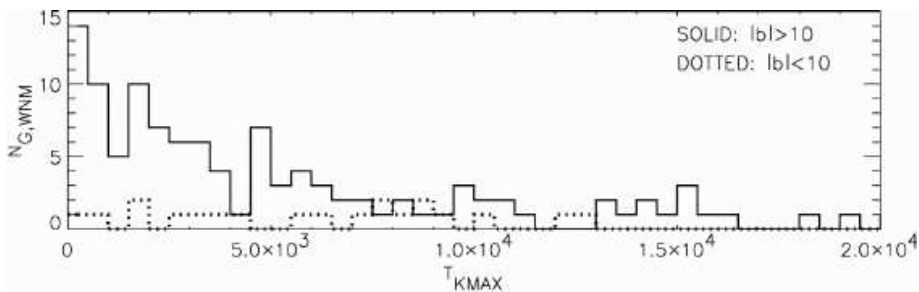}
\caption{Histogram of the WNM kinetic temperature in the case of no non-thermal motions. From Figure 2 in
Heiles \& Troland (2003).}
\label{f:wnm}
\end{center}
\end{figure}

\begin{figure}
\begin{center}
\includegraphics[width=.8\textwidth]{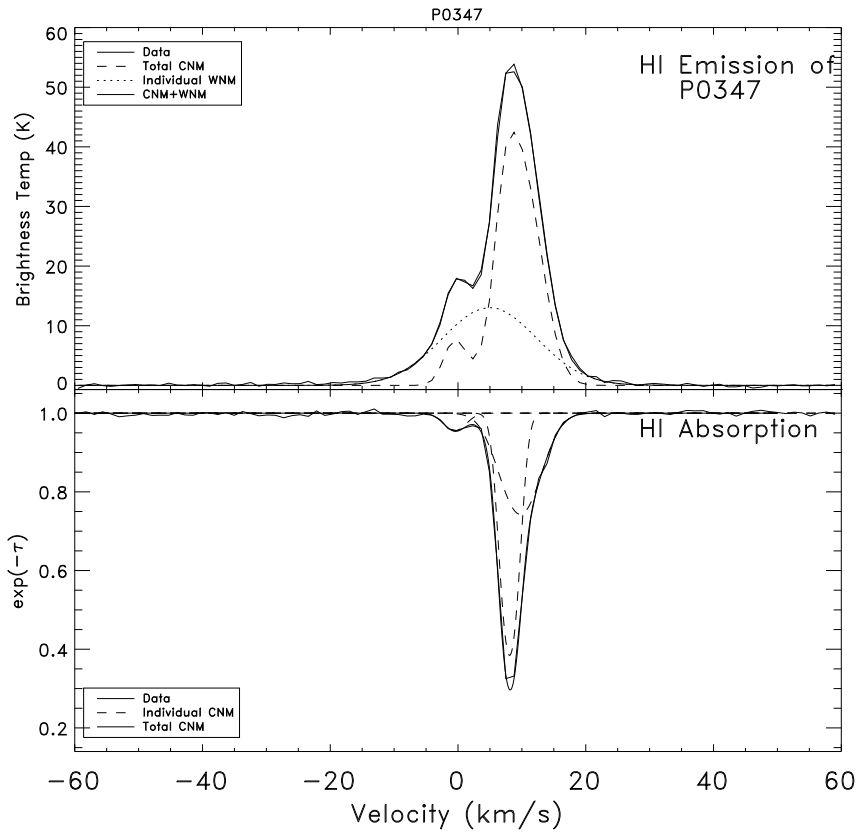}
\vspace{-4cm}
\caption{The HI absorption spectrum (bottom) from our EVLA pilot observations (Begum et al., submitted).
The corresponding HI emission was obtained with the Arecibo radio telescope (top panel).
The absorption spectrum was fit with three CNM Gaussian functions,
shown with the dashed line. The emission spectrum was fit with a 
combination of the CNM (dashed) and the WNM (dotted line) components.}
\label{f:P0347}
\end{center}
\end{figure}

The accretion flows bring fresh star formation fuel to galaxies in various flavors of
diffuse gas which is likely to get integrated with the diffuse interstellar medium (ISM) in the disk.
Again, our home neighborhood (the MW disk) offers a high-resolution view of  the 
crucial physical processes responsible for cycling of interstellar gas across various temperature 
regimes on the way to cold dense clouds, which are considered to be 
precursors of molecular, star-forming entities.

Traditionally, the diffuse neutral ISM is known to exist in two thermal equilibrium states:
the cold neutral medium (CNM) and the warm neutral
medium (WNM; McKee \& Ostriker 1977; Wolfire et al. 2003).
While the CNM properties have been measured extensively, 
surprisingly only  {\it three} direct measurements of the WNM temperature exist thus far. 
The main reason for this observational paucity 
is the low optical depth of the WNM, $\tau \la 10^{-3}$, which creates a 
need for very sensitive radio instruments.
Properties of the WNM are traditionally indirectly inferred only through 
HI emission line profiles.
Out of all ISM phases, the WNM is the least understood, yet it seems to hold the key
to constraining ISM models and the formation of cold interstellar clouds.

One of the key observables that theoretical and numerical models of
the ISM attempt to predict is the gas fraction as a function of temperature.
McKee \& Ostriker (1977) and Wolfire et al. (2003) predict that cold gas should 
dominate, while its enveloping counterpart--the WNM--should be mainly in thermal 
equilibrium and comprise only a few percent of the total diffuse interstellar gas. 
More recent ISM models emphasize the highly dynamic and turbulent character 
of the ISM and the consequences this can have on cold cloud formation.
For example, in Audit \& Hennebelle (2005)'s simulation of a collision of incoming 
turbulent flows a fast condensation of WNM into cold neutral clouds is initiated.
The fraction of cold gas is controlled by turbulence,
and ranges from 10\% in a strong turbulent case to about 30\% in a weak turbulent case. 
Mac Low et al. (2005) simulate how shocks driven 
into warm, magnetized, and turbulent  gas
by supernova explosions create dense, cold clouds. 
They find a continuum of gas temperatures, 
with a fraction of the thermally-unstable WNM ($T<5000$ K) being 
constrained by the star formation rate.

One of the observational studies that has had a large impact on recent 
ISM simulations is the ``Arecibo Millennium" survey of the 21-cm line absorption 
by Heiles \& Troland (2003, HT03). They found that a substantial fraction  (48\%) of the WNM  
is in the thermally unstable phase, with kinetic temperatures in the range of 500$-$5000 K. 
Yet, we must emphasize that HT03 did not measure the WNM directly, but 
inferred its temperature mainly through  observed  narrow HI emission lines (Figure~\ref{f:wnm}).
Two out of three direct measurements of the WNM spin temperature
(Carilli  et al. 1998, Dwarakanath et al. 2002, Kanekar et al. 2003)
also find  $T\sim 3000-4000$ K and support the HT03 results.

To explore possibilities for sensitive HI absorption measurements
with the Expanded Very Large Array (EVLA), we have recently
obtained deep HI absorption spectra  against 
several continuum sources along the lines of sight which have indicated 
the existence of the thermally-unstable WNM with $T<5000$ K  
(Begum et al. 2010). 
As an example, in the direction of source  P0347  HT03 found a narrow 
emission feature at a velocity of 0 \kms~without corresponding 
absorption which indicated a thermally  unstable WNM.  
Figure~\ref{f:P0347} shows our recent 
EVLA absorption spectrum (bottom panel), together with the Arecibo 
HI emission spectrum (top panel),  for this source: 
our detection of an additional weak absorption feature at a velocity 
of $-0.5$ \kms~results  in the best-fit solution without any need for the thermally-unstable WNM.  
This shows that the detection of weak absorption lines, which have been largely missed
in shallow absorption surveys, can
significantly affect the estimated fraction of the thermally-unstable WNM.

Clearly, large samples of very sensitive HI absorption/emission spectra
are needed to characterize the basic properties of the WNM:
temperature, column density, and abundance relative to the CNM.
Current (Westerbork radio telescope and EVLA) and upcoming (ASKAP, MeerKAT, ATA-256) 
radio telescopes are becoming for the first time, technically ready for such experiments.
A dedicated highly sensitive ($\Delta \tau \sim 10^{-4}$) survey of the WNM in absorption is necessary to
measure the basic properties of the WNM and constrain possible scenarios for formation of cold clouds.

In addition, surprisingly little is known about the census and properties 
of cold gas even in very nearby galaxies.
As the ``demography" of cold gas is driven largely by the heating and 
cooling processes -- which rates vary with metallicity, dust-to-gas ratio, and 
the strength of the interstellar radiation field -- significant variations 
of the CNM/WNM properties and abundances are expected
from a theoretical point of view (Wolfire et al. 2003).
The reality is such that, even in our home neighborhood only 
a handful  of measurements exists 
for the cold gas in the SMC and the LMC (Dickey et al.\ 1994, 2000, Marx et al.\ 1997),
typically considered as prototypes of a relatively primitive ISM common in the early Universe.
The only recent attempt to study properties of cold gas in a lower-metallicity 
environment offered by the outskirts of the MW resulted in highly puzzling results.
Dickey et al. (2009) suggest that, contrary to all theoretical predictions, the spin temperature 
of the CNM is constant with Galactocentric radius all the way to 25 kpc.

To be able to study the conversion of cold gas into stars over cosmic
time, we need to start by providing the census of cold gas in nearby galaxies and its 
environmental dependence. 
GASKAP will provide HI absorption spectra for several hundred of radio continuum sources both
behind the Magellanic Clouds and the MW disk to study spatial variations of the
CNM/WNM abundance and their correlations with the underlying physical conditions.
Deeper observations over smaller areas with MeerKAT and ATA-256 should continue this work to 
other nearby galaxies.

\section{What is the origin and nature of interstellar inhomogeneities?}

Interstellar turbulence is an important ingredient
in ISM models and governs many astrophysical processes, including
the cycling across various gas phases, formation and evolution 
of ISM inhomogeneities (McKee \& Ostriker 2007), and the onset of molecule 
formation (Glover et al. 2010).
While pinning down sources of ISM turbulence observationally has been 
hardly explored, detailed numerical simulations of galaxies require 
inclusion of realistic ISM inhomogeneities.
For example, Governato et al. (2010) show how only after tying 
star formation and its feedback to realistic highest-density regions 
can a sufficient removal of the angular momentum be achieved, resulting in reasonable rotation curves.

Statistical studies have proven to be essential in characterizing the inhomogeneous 
and turbulent ISM (Elmegreen \& Scalo 2004, Lazarian 2009). 
However, while  many statistical methods (spatial power spectrum, wavelets, 
probability density functions, principal component analysis etc) have been used,
the interpretation of results is not always straightforward.  
The most challenging issue is the complex relationship between  observables (brightness temperature in
tensity as a function of velocity, in the case of radio observations) and the underlying 
physical quantities (3D density and velocity fields; Lazarian 2009).
In addition, most of these statistical methods require large datasets with a large spatial or 
velocity dynamic range and produce a single, mostly one-dimensional, measure. 
This results in a lack of spatial information about turbulent properties across a given 
interstellar cloud or a galaxy, making a connection with the underlying physical processes
(e.g., presence or absence of star formation, 
strength of magnetic field, presence of shearing motions) very difficult.

Recently, Burkhart et al. (2010) showed that the above problems can be allevaited by using 
modern simulations hand-in-hand with observations.
They developed a new method to provide spatial information about the nature and 
level of interstellar turbulence.
This method is based on applying high-order statistical moments to the HI column density
distribution and bootstrapping the sonic Mach number (${\cal M}_s$) from an 
extensive library of isothermal MHD simulations.
Kowal et al. (2007) used 3D isothermal simulations
of MHD turbulence; their work shows that variance, skewness, and kurtosis
(the 2nd, 3rd and 4th order statistical moments, respectively) 
have a strong dependence on ${\cal M}_s$. 
As the sonic Mach number increases, so does the Gaussian asymmetry of the 
column density PDFs due to gas compression via shocks.  
This implies that the sonic Mach number of turbulence in an interstellar 
cloud can be characterized by applying high-order statistical moments to the
observed column density distribution functions.

\begin{figure}
\hspace{-0.5cm}
\includegraphics[width=1.1\textwidth]{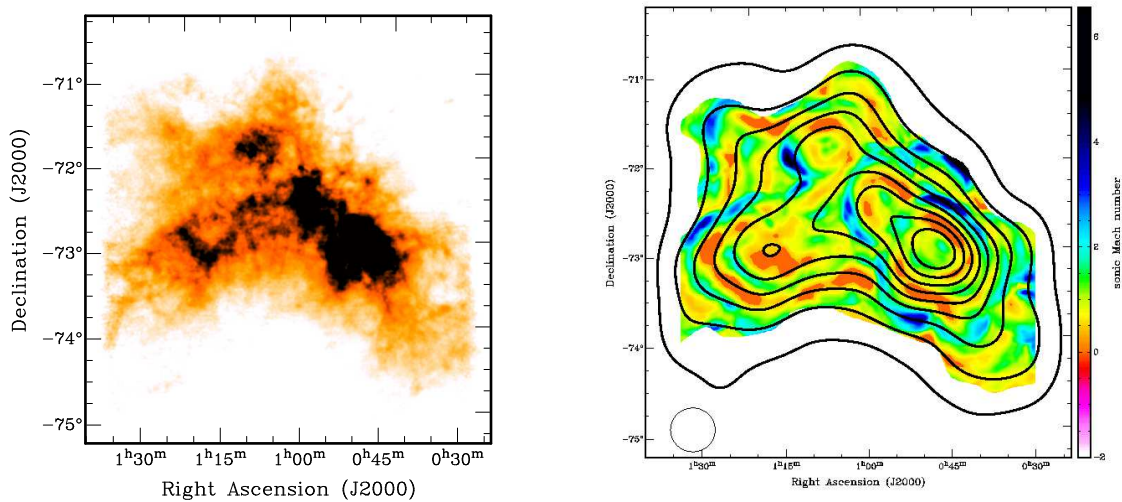}
\hspace{-1cm}
\vspace{-0.5cm}
\caption{(left) The HI column density image of the SMC from Stanimirovic et al. (1999) at 
an angular resolution of  98$''$.
(right) The sonic Mach number image derived from the HI column 
density image of the SMC and overlaid with the HI column density contours 
(from Burkhart et al. 2010). The circle in the bottom-left shows the angular resolution of the image, $\sim30'$.}
\label{f:smc-mach}
\end{figure}

We demonstrated this idea on the HI column density image of the SMC 
(see Figure~\ref{f:smc-mach}, left).
By using the trends provided by simulations, we converted 
high-order statistical images of the HI distribution in the SMC into the sonic Mach number image
shown in Figure~\ref{f:smc-mach}.   This image allowed us, for the first time, to quantify the
fraction of subsonic versus supersonic HI. 
We found that $\sim80$\% of the HI in the SMC is subsonic or
transonic with ${\cal M}_s <2$, while $\sim10$\% appears quiescent with ${\cal M}_s \sim0$.
Another 10\% or so has ${\cal M}_s >2$. The highest supersonic regions,
with ${\cal M}_s \sim 5$, point out large-scale tidal or shearing flows  caused most likely by
the interactions between the SMC, the LMC, and the MW.  
A confluence of observations and numerical simulations
is clearly a powerful way of connecting physical sources and processes with the ISM 
structure formation, which seeds  molecule and later star formation.


However, to have enough data points to reliably calculate
statistical moments we essentially had to smooth the original HI image to
a resolution of 30 arcmin and were therefore not able to reach scales of 
typical HII regions and/or supernovae which are generally considered as the main
turbulence drivers (McCray \& Snow 1979).
The upcoming radio telescopes will ameliorate this problem. 
With an angular resolution of 10$''$-20$''$ provided by ASKAP we will reach linear 
scales of 50-100 pc (at a distance of 60 kpc) for the high-order moment maps.
This will probe the supernova origin of interstellar turbulence at the distance of
Magellanic Clouds. A huge number of high-resolution HI data cubes of 
the Magellanic Clouds, the MW plane (GASKAP) and other nearby galaxies (MeerKAT),
in combination with the high-order moments method,
will sample variations in the nature/level of interstellar turbulence with
varying interstellar environments.
In general, with data volumes increasing by a large factor, an exploration of 
new statistical methods for the analysis of HI data will be even more important in the future.

\section{Conclusions}

While significant effort in the near future will be focused on galaxy evolution by 
observing HI in galaxies at large redshifts, we argue in this paper that 
the understanding of {\it detailed} physical processes that drive evolution of the HI gas
in galaxies is equally important. To expose processes in question at high resolution, 
the nearby Universe offers an unparalleled advantage. 
We have focused here on just three outstanding questions regarding the very first
step in the star-formation cycle in galaxies where upcoming radio telescopes 
(e.g. EVLA, ASKAP, MeerKAT, ATA-256) promise great advances. 

First,  nearby examples of the infalling gaseous tidal tails like the Magellanic Stream 
offer a unique window into how much galaxy halos flavor and tax the 
accretion flows that are postulated to bring fresh star-formation fuel to galaxy disks.
While current analytic and numerical studies suggest highly turbulent environments, 
created by fast shredding of incoming flows by hydrodynamic instabilities, the 
HI clouds in the Stream appear more quiescent with large reservoirs of low column 
density material that is potentially shielding them against destruction.
Second, in our current understanding of the star-formation cycle in galaxies, the WNM
transforms first into the CNM, which further reaches high enough density to form molecules 
and shield them from radiation. Yet, only three direct measurements of the WNM temperature
exist to date in the MW disk. The WNM temperature is a crucial parameter for pinning down
how exactly the warm-to-cold phase transformation occurs in galaxies. Similarly,
very little is known observationally about the abundance of the CNM and 
the CNM/WNM fraction in even nearby galaxies.
Third, interstellar turbulence is a key parameter when modeling the ISM and molecule/star formation,
yet mapping out  turbulent properties across various interstellar environments
and connecting these variations with the underlying energy drivers has been hardly explored.
Statistical methods based on a confluence of observations and numerical simulations
show promising results in this direction and call for higher-resolution data cubes.


The above questions call for extensive, highly-sensitive HI emission and absorption
surveys.  While deep HI emission surveys of tidal tails around galaxies (including the MW) will teach us
about physical properties of accretion flows, deep HI absorption surveys will measure the 
temperature, abundance and interchange of the warm/cold star-formation fuel in galaxies.  
Recent results from Arecibo's GALFA-HI survey, as well as pilot EVLA observations, demonstrate
a high potential of future surveys.
At the same time, to analyze upcoming huge volumes of data,
and to take the data analysis to a higher level, a strong confluence of observations and numerical simulations
is becoming a necessary, not just desirable, approach.

\acknowledgments
It is a great pleasure to thank my collaborators: Lou Nigra, Jay Gallagher, Ayesha Begum, Miller Goss, Carl Heiles, 
Anthony Pavkovich, Patrick Hennebelle, Blakesley Burkhart, Alex Lazarian, and Greg Kowal. 
I also thank the Research Corporation for Science Advancement for their support.


\begin{thebibliography}{39}
 
\bibitem { } Audit, E., Hennebelle, P., 2005, A\&A, 433, 1-13


\bibitem { } Bland-Hawthorn, J., Sutherland, R., Agertz, O., \& Moore, B.\ 2007,
ApJ, 670, L109

\bibitem { } Braun, R., \& Thilker, D.~A.\ 2004, A\&A, 417, 421-435



\bibitem { } Brooks, A.~M., Governato, F., Quinn, T., Brook, C.~B., 
\& Wadsley, J.\ 2009, ApJ, 694, 396 


\bibitem { } Burkhart, B., Stanimirovic, S., Lazarian, A., Kowal, G., 2010,
ApJ, 708, 1204

\bibitem { } Carilli, C. L., Dwarakanath, K. S., Goss, W. M., 1998,
ApJL, 502, L79-L83

\bibitem { } Dickey, J. M., Mebold, U., Marx, M., Amy, S., Haynes, R. F., Wilson, W., 
1994, A\&A, 289, 357-380

\bibitem { } Dickey, J. M., Mebold, U., Stanimirovi{\'c}, S., Stavely-Smith, L., 
2000,  ApJ, 536, 756-772

\bibitem { } Dickey, J. M., Strasser, S., Gaensler, B. M., Haverkorn, M., Kavars, D., 
McClure-Griffiths, N. M., Stil, J., Taylor, A. R., 2009, 
ApJ, 693, 1250-1260

\bibitem { } Dickey, J. M., et al. 2010, in ``The Dynamic ISM: A celebration of the Canadian 
Galactic Plane Survey," ASP Conference Series

\bibitem { } Dwarkanath, K. S., Carilli, C. L., Goss, W. M., 2002, 
ApJ, 567, 940-946

\bibitem { } Elmegreen, B., Scalo, J., 2004, ARAA, 42, 211-273

\bibitem { } Glover, S.C.O.,  Federrath, C., Mac Low, M.-M., Klessen, R. S., 2010,
MNRAS, 404, 2

\bibitem { } Governato, F., Brook, C., Mayer, L., Brooks, A., Rhee, G., Wadsley, J., 
Jonsson, P., Willman, B., Stinson, G., Quinn, T., Madau, P., 2010,
Nature, 463, 203

\bibitem { } Heiles, C., Troland, T. H., 2003, ApJ, 586, 1067-1093


\bibitem { } Heitsch, F., \& Putman, M.~E.\ 2009, ApJ, 698, 1485 

\bibitem { } Hennebelle, P., Audit, E., Miville-Deschenes, M. -A., 2007, 
A\&A, 465, 445-456

\bibitem { } Kalberla, P.~M.~W., \& Haud, U.\ 2006, A\&A, 455, 481 



\bibitem { } Kanekar, N., Subrahmanyan, R., Chengular, J., Safouris, V., 
2003, MNRAS, 346, L57-L61

\bibitem  { } Kere{\v s}, D., Katz, N., Weinberg, D.~H. and Dav{\'e}, R., 2005,
MNRAS, 363, 2

\bibitem { } Kowal, G., Lazarian, A., Beresnyak, A., 2007, ApJ, 658, 423-445

\bibitem Lazarian, A., 2009, Space Science Reviews, 143, 357-385 

\bibitem { } Matthews, D., Staveley-Smith, L., Dyson, P., \& Muller, E.\ 2009, ApJL, 691, L115 

\bibitem{ } McCray, R., \& Snow, T.~P., Jr.\ 1979, ARA\&A, 17, 213 


\bibitem { } McKee, C. F., Ostriker, J. P., 1977,  ApJ, 218, 148-169

\bibitem { } McKee, C. F., Ostriker, E.C., 2007, ARA\&A, 45, 565

\bibitem { } Mac Low, M., Balsara, D., Kim, J., de Avillez, M., 
2005, ApJ, 626, 864-876

\bibitem { } Marx, M., Dickey, J. M., Mebold, U., 1997, A\&ASS, 126, 325-334

\bibitem { } Mori, M., \& Burkert, A.\ 2000, ApJ, 538, 559

\bibitem { } Murray, S.~D., White, 
S.~D.~M., Blondin, J.~M., \& Lin, D.~N.~C.\ 1993, ApJ, 407, 588 

\bibitem { } Peek, J.E.G. \& Heiles, C., 2008, astro-ph/0810.1283

\bibitem { } Putman, M. E. et al., 2009, Astro2010: The Astronomy and Astrophysics Decadal 
Survey, Science White Papers, no. 241

\bibitem { } Quilis, V. \& Moore, B.\ 2001, ApJL, 555, L95-L98

\bibitem { } Shull, J.~M., Jones, 
J.~R., Danforth, C.~W., \& Collins, J.~A.\ 2009, ApJ, 699, 754 


\bibitem { } Stanimirovi{\'c}, S., Staveley-Smith, L., Dickey, J. M., 
Sault, R. J., Snowden, S. L., 1999, MNRAS, 302, 417-436



\bibitem { } Stanimirovi{\'c}, S., et al. \ 2006, ApJ, 653, 1210-1225


\bibitem { } Stanimirovi{\'c}, S., Hoffman, S., Heiles, C., Douglas, K., 
Putman, M., Peek, J., 2008,  ApJ, 680, 276-286

\bibitem { } Vieser, W., \& Hensler, G.\ 2007, A\&A, 472, 141 

\bibitem { } Wolfire, M. G., McKee, C. F., Hollenbach, D., 
Tielens, A. G. G. M., 2003, ApJ, 587,  278-311



\end{thebibliography}
\end{document}